\documentclass[aps,prl,twocolumn,showpacs,groupedaddress]{revtex4}

\usepackage{graphicx}
\usepackage{dcolumn} 
\usepackage{bm}
\usepackage{amssymb} 

\begin{document}

\title{Observation of Periodic $\pi$-Phase Shifts in
Ferromagnet-Superconductor Multilayers}

\author{V.~Shelukhin$^{1}$, A.~Tsukernik$^{2}$, M.~Karpovski$^{1}$, Y.~Blum$^{1}$,
K.B.~Efetov$^{3}$, \\A.F.~Volkov$^{3}$,
T.~Champel$^4$, M.~Eschrig$^{4}$, T.~L\"{o}fwander$^{4}$, G.~Sch\"{o}n$^{4}$,
and A.~Palevski$^{1}$}

\affiliation{
$^{1}$School of Physics and Astronomy,
Raymond and Beverly Sackler Faculty of Exact Sciences,
Tel Aviv University, Tel Aviv 69978, Israel\\
$^{2}$University Research Institute for Nanoscience and
Nanotechnology, Tel Aviv University, Tel Aviv 69978, Israel\\
$^{3}$ Theoretishe Physik III, Ruhr-Universit{\"a}t Bochum, Germany\\
$^{4}$ Institut f{\"u}r Theoretishe Festk{\"o}rperphysik, Universit{\"a}t Karlsuhe,
76128 Karlsruhe, Germany}

\date{\today}

\begin{abstract}
We report complementary studies of the critical temperature and the
critical current in ferromagnet (Ni) - superconductor (Nb)
multilayers. The observed oscillatory behavior of both quantities upon
variation of the thickness of the ferromagnetic layer is found to be
in good agreement with theory. The length scale of oscillations is
identical for both quantities and is set by the magnetic length corresponding to
an exchange field of 200 meV in Ni. The consistency between the
behavior of the two quantities provides strong evidence for periodic
$\pi$ phase shifts in these devices.
\end{abstract}


\maketitle

Devices of superconducting materials as for example Josephson
junctions \cite{Golubov}, have proven exceptionally useful in many
fields of physics. In bulk conventional superconductors, the spin
degree of freedom is frozen out by the spin-singlet Cooper pair
formation. The proximity to a ferromagnetic material \cite{Buzdin},
however, opens up the spin as an additional degree of freedom
\cite{Bergeret,Eschrig}.  Consequently, the construction of Josephson
junctions of superconductor-ferromagnet-superconductor (S-F-S)
materials has attracted considerable attention recently within the
emerging field of spintronics \cite{Prinz}.

One intriguing effect associated with the new spin degree of freedom
in an S-F-S Josephson junction is the thermodynamic stability of a
phase difference $\pi$ between the two superconductors for certain
parameter ranges of the middle ferromagnetic material
\cite{Bulaevskii}. The $\pi$ phase is a result of a peculiar
superconducting proximity effect in the ferromagnet (F). The two spin
species are split in energy by the exchange field $E_{ex}$, which
leads to an oscillatory behavior of the proximity induced pair
amplitude in the ferromagnet.  As a result, properties such as the
critical temperature $T_c$ and the critical current $I_c$ are
non-monotonic, oscillating and decaying functions of increasing
ferromagnetic thickness.  For the Josephson junction, the free energy
loss due to the energy splitting can be compensated for by a
spontaneous appearance of a superconducting phase difference of $\pi$
over the junction. This additional degree of freedom leads to a series
of $0\rightarrow\pi$ and $\pi\rightarrow 0$ transitions, that can be
observed as zero crossings of $I_c$ and as kinks and minima of $T_c$
with varying ferromagnetic layer thickness \cite{Radovic}.

Previous studies of these effects have been focused on ferromagnetic
alloys sandwiched between two superconductors, because the
corresponding oscillation wavelength is quite long and is easily
resolved \cite{Ryazanov,Kontos}. It is important for the development
of applications to also understand devices including strong
ferromagnets as iron, cobalt, or nickel, but the short oscillation
wavelength in these materials is harder to resolve. In previous
reports on devices made of strong ferromagnets either the $T_c$
variations \cite{Jiang,Sidorenko} or the $I_c$ variations
\cite{Palevski,Obi,Bell} were considered as function of ferromagnetic
layer thickness. However, sample preparation techniques make a direct
comparison of the different experiments difficult. For a successful
theoretical understanding a high control of material properties is
required.

In this Letter we report studies of both the critical current and the
critical temperature variations as function of the F layer thickness
in S-F-S junctions made of one set of materials, namely Nb-Ni-Nb
junctions prepared under identical conditions. We find that both
quantities, $I_c$ and $T_c$, vary on the same scale, the magnetic
length $L_M=\sqrt{\hbar D_F/E_{ex}}$ set by the properties of the
ferromagnet only ($D_F$ is the diffusion constant in the
ferromagnet). By a detailed comparison of our measurements with
theory, we find consistent fits for an exchange field of
$E_{ex}=200\,\text{meV}$ in Ni.

The samples for the $I_c$ measurements had a
$10\times10\,\mu\text{m}^{2}$ cross sectional area and were
fabricated with a standard photolithography technique. The process
contained three stages of lithography: liftoff of the bottom Nb/Cu
layers, liftoff of the variable thickness Ni layer, and liftoff of
the top Cu/Nb layers. We fabricated two sets of nine
Nb-Cu(Au)-Ni-Cu(Au)-Nb junctions with variable Ni thickness in the
range of 35~{\AA} to 75~{\AA} in steps of 5~{\AA} (set I) and
4~{\AA} (set II). The thickness of each Nb layer is 2000{~\AA},
while the total thickness of the Cu is 2400$\pm$250{~\AA} (Au -
500$\pm$50{~\AA}). We show the layout of the junctions for the $I_c$
measurements in Fig.1

\begin{figure}[t]
\centering
\includegraphics[width=\linewidth,height=2.4in]{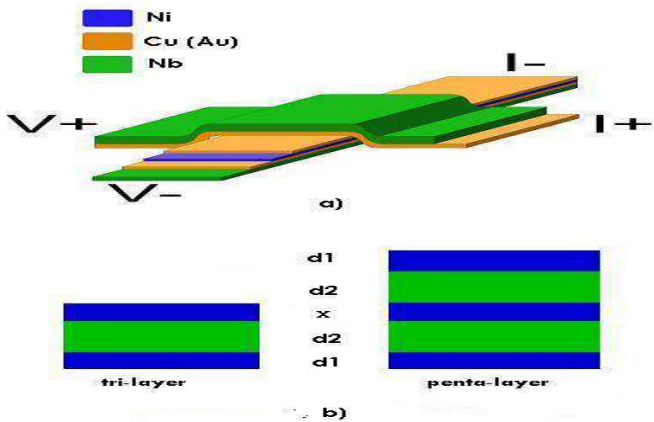}
\caption{Schematic layout: a) of the junctions for $I_c$
experiments,
                           b) of planar samples for $T_{c}$ experiments.}
\label{fig:1}
\end{figure}

The Nb films were sputtered using a magnetron gun and were covered in
situ with a Cu (set I) or a Au (set II) layer by thermal evaporation
to prevent Nb oxidation. The ferromagnet layers of Ni were e-gun
evaporated in a separate vacuum chamber at a pressure of
$2\cdot10^{-7}$ torr and subsequently covered in situ by Cu or Au.
The variation of the Ni thickness was achieved by a specially designed
shutter, which exposed the samples in sequence, so that every sample
was exposed to the evaporating Ni for additional fragments of
time. Because all samples within one set were prepared simultaneously,
all layer interfaces are nominally identical and the only difference
between the samples is their Ni thickness. The critical current was
measured by passing a DC current with a small AC modulation through
the sample. The AC voltage, which appeared above the critical DC
current, was picked up by a lock-in amplifier operated in a
transformer mode. The measurements were performed in a $^{4}$He
cryostat in the range from 4.2~K down to 1.5~K.

The samples for the $T_{c}$ studies were prepared by an in-situ
evaporation of Ni and Nb layers without photolithography, see
Fig.~1b. Two sets of structures, each containing 16 samples were
fabricated. The first set of samples contains only a single layer of
Nb and was obtained by sequential deposition of
Ni(30{\AA})-Nb(430{\AA})-Ni(x), where Ni(x) was varied from 0 to
37~{\AA} in steps of $\sim$2.5~{\AA}. The second set contains two Nb
layers, namely
Ni(30{\AA})-Nb(430{\AA})-Ni(x)-Nb(430{\AA})-Ni(30{\AA}), and was
prepared in a similar manner. The thicknesses of the bottom and top Ni
layers, as well as the thicknesses of the Nb layers, were chosen such
that the bulk $T_{c}$ of Nb was suppressed. This increases the
sensitivity of $T_{c}$ to variations of the thickness of the center Ni
layer.

For strong ferromagnets like Ni, the magnetic length $L_M$ is below
20~{\AA}. The first $0\rightarrow\pi$ transition is therefore expected
to occur for very thin films. However, Ni films thinner than a few
tens of {\AA}ngstr\"oms prepared by standard e-gun evaporation is not
expected to be homogeneous or to perfectly cover a metallic
surface. This is a problem for measurements of $I_c$ in S-F-S
junctions with a very thin F-layer, since uncovered regions
short-circuit the junction. For thermodynamic measurements, such as
measurements of $T_c$, an inhomogeneous coverage is less of a
problem. At the same time, the effect we address decays exponentially
with the layer thickness, and it is undesirable to have very thick
films, in particular for thermodynamics measurements. With this in
mind, we have chosen Ni-film thicknesses ranging from 0 to 35~{\AA}
for the $T_c$ measurements and thicknesses ranging from 35 to 75~{\AA}
for the $I_c$ measurements.

The results of the critical current measurements are shown in Fig.~2.
We have also included the set published earlier in
Ref.~\cite{Palevski}.
\begin{figure}[b]
\centering
\includegraphics[width=1\linewidth,height=2.8in]{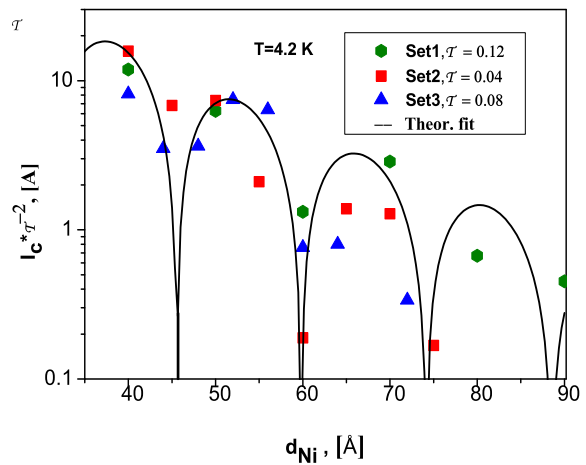}
\caption{The critical Josephson current $I_c$ as function of the
Ni-layer thickness $d_{\sf Ni}$. The solid line shows the
theoretical curve with $E_{ex}=200$~meV used as a fitting parameter.
The dots are the experimental data.} \label{fig:2}
\end{figure}
Since the sets were prepared separately and are somewhat different in
their normal metal constituents, they are expected to have different
interface properties reflected as a different transmission coefficient
${\cal T}$ of the superconductor-ferromagnet interface, which enters
the calculations of $I_c$ as a prefactor ${\cal T}^2$
\cite{Efetov}. Therefore, we have normalized the values of $I_c$ by
${\cal T}^2$ varying between the sets, as indicated in the legend of
Fig.~2. We use the Landauer formula to estimate a lower limit for
${\cal T}$, under the assumption that the entire resistance of the
junction arises from the two S-F interfaces. By using a typical
resistance value $\text{R}=200~\mu\Omega$ for our junctions we get
${\cal T}_{min}\sim 0.012$. The values of ${\cal T}$ used for all of
our sets are indeed larger than the estimated lower limit.
We emphasize that irrespective of which theory our experimental data
is compared with, every $\approx 15$~{\AA} of Ni the phase over the
Nb-Ni-Nb junction changes from 0 to $\pi$ (or $\pi$ to 0)
\cite{Palevski_note}. This implies that we should expect the first
pronounced minimum in the variation of $T_{c}$ at a Ni layer thickness
of $d_{\sf min} \approx 15 $~{\AA }.

The theoretical curve for the critical current at $T$=4.2~K normalized
by ${\cal T}^{2}$ (solid line in Fig.~2) was calculated with the
theory of Ref.~\cite{Efetov} using the following parameters: the Fermi
velocity $\upsilon_{F}=2.8\cdot 10^{7}$~cm/s \cite{Petrovykh}, a
critical temperature of Nb $T_c=8.5$~K, the exchange energy
$E_{ex}=200$~meV, and the mean free path $\ell=28${~\AA}. The magnetic
length in terms of these parameters can be determined to be
$L_M=10$~{\AA}. Our result for the exchange energy is somewhat higher
than the results obtained from spin-resolved photoemission
spectroscopy where the splitting between spin-up and spin-down bands
at the Fermi energy ranged between $2E_{ex}=200$~meV and 350~meV
\cite{Heiman}. The theoretical predictions for 2$E_{ex}$ are typically
higher and range from 600~-~850~meV \cite{Dietz}. Our value of
$2E_{ex}=400$~meV falls in between the above experimental and
theoretical values, and is consistent with the values in
Ref.~\cite{Sidorenko} of $E_{ex}=220$~meV (corresponding to a magnetic
length of 8.8{~\AA}), obtained from measurements of $T_{c}$ in Ni-Nb
bi-layers.

The theoretical prediction of the minima in Fig.~2 were obtained
with Eq.~(19) in Ref.~\cite{Efetov}. The oscillation period for
large thicknesses is thus not equal to the thickness where the first
minimum in $I_c$ is predicted by the theory. This is due to the fact
that the first minimum occurs at a thickness smaller than the mean
free path $\ell$. Equal spacing of the minima only takes place in
the regime $d>\ell$. The theoretical prediction for the first
minimum using the above fit is $d_{\sf min}\approx 17$~{\AA}.

It was noted recently that the critical current of junctions with a
given thickness of the ferromagnet Py, scatter considerably
\cite{Bell}. We confirm this effect in our Ni junctions.  We believe
that this phenomenon is related to the domain structure of the
ferromagnet. It was recently shown that variations in the domain
configuration lead to considerable variations in $I_c$, provided that
the magnetic flux through typical domains is of the order of the flux
quantum $\Phi _{0}$ \cite{Volkov}. By using a magnetization for Ni of
$M_{s}=500$~Oe, we estimate for a 50~{\AA}~$\times$~1$\mu$m domain
cross section a flux of approximately one flux quantum. In order to
average the influence of domains, we have measured different junctions
with the same thickness of the ferromagnet layer.

Fig.~3 shows the variation of $T_{c}$ in the Ni-Nb-Ni(x)-Nb-Ni
multilayer structure versus the thickness of the Ni(x) layer. The data
contains a pronounced minimum around $x_{\sf min}$=17{~\AA}, which is
in excellent agreement with the expected $0\rightarrow\pi$ transition
at $d_{\sf min}\approx $15-17{~\AA} implied by the critical current
measurements above. In order to ensure that the observed minimum
arises from a $0\rightarrow\pi$ transition, the variation of $T_{c}$
in a Ni-Nb-Ni(x) structure was measured for the same range of Ni(x)
thicknesses, see Fig.~3. For symmetry reasons, $T_c$ should vary twice
as fast for the trilayer compared to the $0$-phase of the
pentalayer. The absence of a pronounced local minimum in the trilayer
with a single Nb layer therefore undoubtedly indicates that the
minimum observed for the pentalayer containing two Nb layers must
arise from a $0\rightarrow\pi$ transition. The kink-like change of
$T_c$ at 17{~\AA} also suggest that there is a crossing point of two
phases, namely the $0-$ and $\pi$-phases.

\begin{figure}[t]
\centering
\includegraphics[width=1\linewidth,height=2.8in]{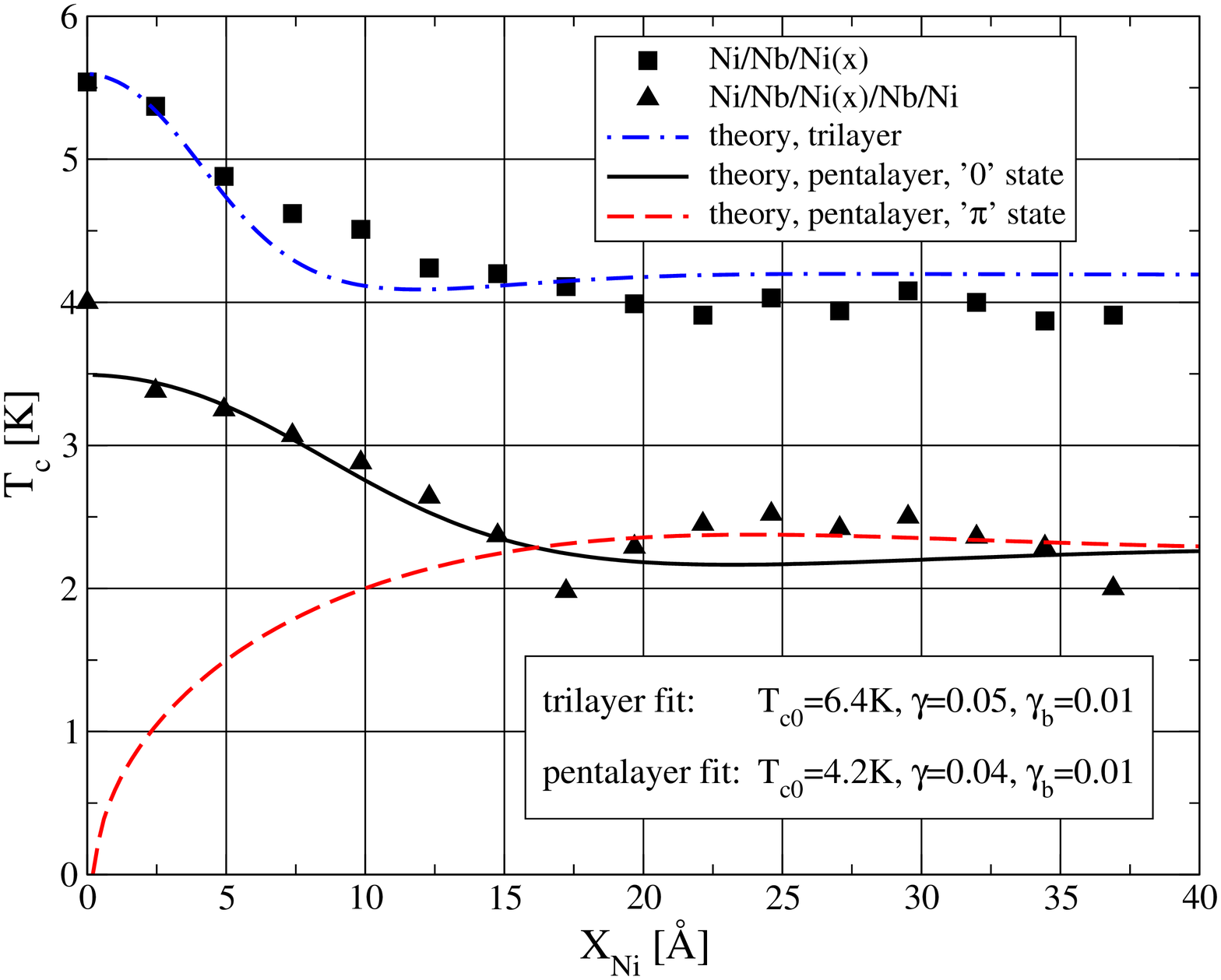}
\caption{Critical temperature $T_c$ as a function of Ni-layer
thickness $x_{\sf Ni}$ for the trilayer and pentalayer structures.
The experimental data are shown as symbols, while the curves are fits
to the data with the theory discussed in the text. For the pentalayer,
the highest $T_c$ is obtained for a zero junction (full line) for
$d_{\sf Ni}<$16~{\AA}, and for a $\pi$-junction (dashed line) in the
range 16~{\AA}$<d_{\sf Ni}<$40~{\AA}.}
\label{fig:3}
\end{figure}

In order to compare our experimental results for $T_{c}$ with theory
we solve the gap equation and compute $T_c$ of our systems with the
quasiclassical Green's function technique in the diffusive
approximation \cite{usadel}. Near $T_c$, the order parameter
$\Delta\ll T_c$ and the Usadel equation can be linearized. We have
generalized the results for symmetric trilayers by Fominov {\it et
al.} in Ref.~\cite{fominov_trilayer} to asymmetric F$_1$-S-F$_2$
trilayers and symmetric F$_2$-S-F$_1$-S-F$_2$ pentalayers. Instead
of discretizing the spatial coordinate we Fourier-series expand the
order parameter and find $T_c$ by studying the resulting eigenvalues
of the gap equation.  With this technique \cite{details}, the accuracy
as well as the speed of the numerics are improved immensely compared
with previously used methods \cite{fominov_trilayer,Champel}.

Several material parameters serve as input to the model: the exchange
field $E_{ex}$ of Ni, the critical temperature $T_{c0}$ of Nb in the
absence of the Ni layers, and the diffusion constants of Ni ($D_F$)
and Nb ($D_S$). The boundary conditions \cite{KupLuk} at the Ni-Nb
interface are expressed in terms of a normalized boundary resistance
$\gamma_b$ and the conductivity mismatch $\gamma$ between the Ni and
Nb materials. We have considered these two quantities as free
parameters.

Our fits of the experimental data for $T_c$ as function of Ni layer
thickness with the above theory are shown as curves in Fig.~3. We use
as input parameters the exchange field {$E_{ex}=200$ meV} and the
diffusion constants {$D_F=2.8$~cm$^2$/s} and {$D_s=3.9$~cm$^2$/s}
(with $D=\frac{1}{3}\upsilon_f\ell$) obtained from the fit of $I_c$ in
Fig.~2. The fit parameters are the bulk Nb transition temperature
$T_{c0}$, the interface resistance $\gamma_b$, and the materials'
conductivity mismatch $\gamma$. The fits indicate that $T_{c0}$ of the
batches of trilayers and pentalayers differ, while other sample
characteristics remained essentially the same. Although all samples
within the trilayer set and within the pentalayer set were prepared in
situ, both sets were evaporated separately. We assign the difference
in $T_{c0}$ to this fact.

We have calculated $T_c $ as a function of $x_{\sf Ni}$ for the
pentalayer for zero phase difference and for $\pi$ phase difference
between the two superconductors, using the {\it same} parameters.  The
corresponding curves are shown in Fig.~3 as full and dashed lines. The
$0\rightarrow\pi$ transition takes place where the two $T_c$ curves
cross. We note, that the fit parameter $T_{c0}$ is determined by the
small thickness data points, and the remaining fit parameters $\gamma$
and $\gamma_b$ are determined by the fitting of the zero-phase
curve. Having no additional fit parameter, the $0\rightarrow\pi$
crossing at $x_{\sf Ni}=16 $~{\AA} is in remarkable agreement with the
experimental data for $T_c$, and with the prediction of the $I_c$ data
fit. We also fit the experimental data for the trilayer with very
similar interface parameters.

We would like to mention that we expect corrections to the Usadel
theory when the exchange field is large. From the fit of the $I_c$
data, where such corrections were taken into account, we see that
$\ell\sim L_M$, while Usadel theory works well for $\ell \ll L_M$.
Nevertheless, we obtain a remarkably good fit for $T_c$ as function
of $x_{\sf Ni}$.
This is probably due to the fact that the Ni-layers for the $T_c$ measurements
are quite thin, in which case surface disorder is relevant and
justifies the use of Usadel theory. We note that the surfaces are
characterized in this case by strong disorder with
a large number of point contacts (high-transmission channels).
This is consistent with our observation of regions with short-cuts in the samples
used for the $T_c$ measurements, which prevented us
from extending our $I_c$ measurements to $d_{\sf Ni}<30 $~{\AA}.

In summary we have demonstrated that both the critical Josephson
current and the critical temperature of Nb-Ni multilayers vary with
the Ni thickness with approximately the same period, 16$\pm$1{~\AA}.
We deduce from the period a magnetic length $L_{M}$=10{~\AA},
corresponding to an exchange energy of $E_{ex}$=200~meV.  By measuring
$T_c$ in Ni-Nb pentalayers, we have observed a $0\rightarrow\pi$
transition at a thickness for the central Ni layer consistent with the
theoretical prediction using Usadel theory.  For higher thicknesses,
we see further $\pi\rightarrow 0$ and $0\rightarrow\pi$ transitions in
the critical Josephson current, consistent with the period in our
$T_c$ measurements and with the predictions of theory.  Our results
demonstrate the feasibility of using strong ferromagnetic materials in
the design of Josephson devices for future applications.

This research was supported by the German-Israeli Foundation for
Scientific Research and Development, and by the Israel Science
Foundation. A.F.V. and K.B.E. would like to thank SFB 491
for financial support.
In addition we acknowledge support from the Deutsche
Forschungsgemeinschaft within the Center for Functional Nanostructures
(T.C., M.E., and G.S.), and the Alexander von Humboldt Foundation (T.L.).


\begin{thebibliography}{10}
\bibitem{Palevski_note}
The observed periodicity of $I_{c}$ oscillations in Fig.~\ref{fig:2} is about
half the periodicity that was deduced from the previously published
data alone \cite{Palevski}, where the variation of the Ni thickness
was made in steps of 10{~\AA}.  With the current data we are able to
resolve all minima in the $I_{c}$ curves as function of layer
thickness.
\bibitem{details} The details will be published elsewhere.
\end{thebibliography}
\end{document}